\documentclass[journal]{IEEEtran}
\ifCLASSINFOpdf
\else
\fi

\usepackage{graphicx}

\usepackage{float}
\usepackage{multirow}
\usepackage{lineno}

\usepackage{hepunits}

\usepackage{cite}

\usepackage{authblk}

\begin{document}

\begin{titlepage}

\thispagestyle{empty}
\def\thefootnote{\fnsymbol{footnote}}       

\begin{center}
\mbox{ }

\end{center}
\begin{center}
\vskip 1.0cm
{\Huge\bf
Precision Luminosity of LHC \mbox{Proton-Proton} Collisions 
at 13\,TeV Using Hit-Counting with TPX Pixel Devices
}
\vskip 1cm
{\LARGE\bf 
Andr\'e Sopczak$^1$,
Babar Ali$^1$,
Thanawat Asawatavonvanich$^1$, \\
Jakub Begera$^1$,
Benedikt Bergmann$^1$,
Thomas Billoud$^2$,
Petr Burian$^1$,
Ivan Caicedo$^1$,
Davide~Caforio$^1$,
Erik Heijne$^1$,
Josef Jane\v cek$^1$,\\
Claude Leroy$^2$,
Petr M\' anek$^1$,
Kazuya Mochizuki$^2$,
Yesid Mora$^1$,
Josef Pac\' ik$^1$,
Costa Papadatos$^2$,
Michal Platkevi\v c$^1$,
\v St\v ep\' an Polansk\' y$^1$,
Stanislav Posp\'i\v sil$^1$,
Michal Suk$^1$,
Zden\v ek Svoboda$^1$
\bigskip
}

\Large 
$^1$Institute of Experimental and Applied Physics, 
Czech Technical University in Prague, Czech Republic\\
$^2$Group of Particle Physics, University of Montreal, Canada 

\vskip 1.0cm
\centerline{\Large \bf Abstract}
\end{center}

\vskip 1.2cm
\hspace*{-0.5cm}
\begin{picture}(0.001,0.001)(0,0)
\put(,0){
\begin{minipage}{\textwidth}
\Large
\renewcommand{\baselinestretch} {1.2}
A network of Timepix (TPX) devices installed in the ATLAS cavern measures the LHC 
luminosity as a function of time as a stand-alone system.
The data were recorded from 13\,TeV proton-proton collisions in 2015.
Using two TPX devices, the number of hits created by particles passing 
the pixel matrices was counted.
A van der Meer scan of the LHC beams was analysed using bunch-integrated luminosity 
averages over the different bunch profiles for an approximate absolute luminosity normalization.
It is demonstrated that the TPX network has the capability to measure 
the reduction of LHC luminosity with precision.
Comparative studies were performed among four sensors (two sensors in each TPX device)
and the relative short-term precision of the luminosity measurement
was determined to be 0.1\% for 10\,s time intervals.
The internal long-term time stability of the measurements was below 0.5\% for 
the data-taking period.
\renewcommand{\baselinestretch} {1.}

\normalsize
\vspace{1.5cm}
\begin{center}
{\sl \large
Presented at the IEEE 2016 Nuclear Science Symposium, Strasbourg, France 
\vspace{-6cm}
}
\end{center}
\end{minipage}
}
\end{picture}
\vfill

\end{titlepage}

\newpage
\setcounter{page}{1}

\title{Precision\,Luminosity\,of\,LHC\,\mbox{Proton-Proton}\,Collisions 
       at\,13\,TeV\,Using\,Hit-Counting\,with\,TPX\,Pixel\,Devices}

\author{Andr\'e Sopczak, {\em Senior Member, IEEE},
Babar Ali,
Thanawat Asawatavonvanich,
Jakub Begera,
Benedikt Bergmann,
Thomas Billoud,
Petr Burian,
Ivan Caicedo,
Davide~Caforio,
Erik Heijne, {\em Fellow IEEE},
Josef Jane\v cek,
Claude Leroy, {\em Member IEEE},
Petr M\' anek,
Kazuya Mochizuki,
Yesid Mora,
Josef Pac\' ik,
Costa Papadatos,
Michal Platkevi\v c,
\v St\v ep\' an Polansk\' y,
Stanislav Posp\'i\v sil, {\em Senior Member, IEEE},
Michal Suk,
Zden\v ek Svoboda \vspace*{-6mm}
\thanks{A. Sopczak, 
B.~Ali,
T.~Asawatavonvanich (IAESTE student),
J.~Begera,
B.~Bergmann,
P.~Burian,
I.~Caicedo,
D.~Caforio,
E.~Heijne,
J.~Jane\v cek,
P.~M\' anek,
Y. Mora,
J.~Pac\' ik,
M.~Platkevi\v c,
S.~Polansk\' y,
S.~Posp\'i\v sil,
M.~Suk, and
Z.~Svoboda
are with the 
Institute of Experimental and Applied Physics, 
Czech Technical University in Prague, Horska 3a/22, CZ-128\,00, 
Czech Republic (e-mail: andre.sopczak@cern.ch).}
\thanks{
T.~Billoud,
C. Leroy, 
K. Mochizuki, and
C.~Papadatos
are with the 
Group of Particle Physics, University of Montreal, 
  2900 boul. \'Edouard-Montpetit, Montr\'eal QC  H3T 1J4T, Canada.}
}

\maketitle

\thispagestyle{empty}

\begin{abstract}
A network of Timepix (TPX) devices installed in the ATLAS cavern measures the LHC 
luminosity as a function of time as a stand-alone system.
The data were recorded from 13\,TeV proton-proton collisions in 2015.
Using two TPX devices, the number of hits created by particles passing 
the pixel matrices was counted.
A van der Meer scan of the LHC beams was analysed using bunch-integrated luminosity 
averages over the different bunch profiles for an approximate absolute luminosity normalization.
It is demonstrated that the TPX network has the capability to measure 
the reduction of LHC luminosity with precision.
Comparative studies were performed among four sensors (two sensors in each TPX device)
and the relative short-term precision of the luminosity measurement
was determined to be 0.1\% for 10\,s time intervals.
The internal long-term time stability of the measurements was below 0.5\% for 
the data-taking period.
\end{abstract}

\vspace*{-6mm}

\section{Introduction}
\label{sec:introduction}

A TPX detector network~\cite{claude} of sixteen devices was installed in the ATLAS
cavern at CERN. Each TPX device consists of two stacked hybrid silicon 
pixel sensors. The silicon sensors have a matrix of $256\times256$ pixels 
of $55\,\micron$ pitch, and thickness of $300\,\micron$ (further indicated as layer-1)
and $500\,\micron$ (layer-2)~\cite{benedikt}.
The readout chips connected to these sensors have the original Timepix 
design~\cite{llopart,llopart2}.
The installation of the TPX devices took place during the LHC shutdown 
transition from Run-1 to \mbox{Run-2} in 2013/2014.
These double-layer TPX devices replaced the previously operational network 
that employed single-layer Medipix (MPX) 
assemblies~\cite{previous, analysisRadiaField:2013}.

These devices measure the primary and secondary particle fluxes resulting 
from 13~TeV proton-proton collisions.
The data were taken in 2015 during the first year of LHC Run-2 operation.
Precision luminosity measurements are of particular importance for many physics 
analyses in high-energy physics.

The use of the TPX network for luminosity measurements has several advantages 
compared to the previous luminosity measurements~\cite{Sopczak:2015qqj} at 
LHC during Run-1 that used MPX devices.
Regarding the luminosity monitoring, the two-layer hodoscope structure 
of the TPX devices doubles the measurement statistics and 
allows one to determine the precision and long-term time stability of individual TPX devices.
The dead-time caused by the readout was reduced from about 6\,s to 0.12\,s allowing 
a much higher data acquisition rate.
Also, the TPX devices are operated in three different modes~\cite{llopart,llopart2}:
hit-counting,
time-over-threshold (energy deposits and cluster-counting), and
time-of-arrival (cluster-counting).
Furthermore, 
the proton-proton collision energy increased from 8~TeV at LHC during Run-1 to 13~TeV at Run-2 
which opened a new energy frontier for luminosity measurements in colliders.

The TPX network is self-sufficient for luminosity monitoring.
It collects data independently of the ATLAS data-recording chain, 
and provides independent measurements of the bunch-integrated LHC luminosity.
In particular, van der Meer (vdM) scans~\cite{vdm} can be used for absolute 
luminosity calibration.

The detection of charged particles in the TPX devices is based on 
the ionization energy deposited by particles passing through the silicon sensor. 
The signals are processed and digitized during an adjustable 
exposure time
(frame acquisition time) for each pixel. 
Neutral particles, namely neutrons, however, 
need to be converted to charged particles before they can be detected. 
Therefore, a part of each silicon sensor is covered by $^6$LiF and polyethylene
converters~\cite{vykydal,benedikt}.

Thirteen out of the sixteen installed devices have been used for the luminosity analysis.
Two devices were found to be inoperational after the closing of the ATLAS detector and
one device was intentionally located far away from the interaction point 
and therefore it was unusable for luminosity measurements.
Table~\ref{tab:tpx_detectors} lists the locations of the devices TPX02 and TPX12 used in this 
analysis, and their numbers of registered passing particles (clusters).
It is noted that the number of clusters for the $500\,\micron$ sensor is 
about 20\% larger compared to the $300\,\micron$ sensor.
This percentage is lower than might be expected from the 40\% larger sensitive volume, 
because the extended clusters induced by a single particle count only `one' 
in the thin as well as in the thick sensor. 
The number of photon conversions and fast neutron interactions, 
however, will increase with the sensitive volume.
The analysis described in this article is focused on the precision luminosity determination with 
the devices TPX02 and TPX12, which are operated with 1\,s exposure time 
and analysed in hit-counting mode.
As their positions are very similar in $R$ and $Z$ coordinates on opposite sides of the 
proton-proton interaction point, their count rates are very similar.
During 2015 LHC proton-proton collisions typical luminosities were
$\rm {\cal L} = $ 3-5$\rm \cdot10^{33}\,cm^{-2}s^{-1}=$ 3000-5000$\rm \,\micro b^{-1}s^{-1}$ and
the TPX count rate was a few $10^6$ hits/s per frame.

\begin{table}[bp]
\caption{TPX device locations with respect to the interaction point.
$Z$ is the longitudinal distance from the interaction point and
$R$ is the distance from the beam axis. Their uncertainty is about 10\,mm.
\vspace*{-4mm}
}
\label{tab:tpx_detectors}
\begin{center}
\renewcommand{\arraystretch}{1.2}
             \begin{tabular}{crrcc}
                        \hline\hline
                Device  & $Z$~~  & $R$~~  & \multicolumn{2}{c}{TPX clusters per unit sensor area and}   \\
                        & (mm) & (mm) & \multicolumn{2}{c}{per unit luminosity $(\rm cm^{-2}/ nb^{-1})$} \\
                        &     &     &  \hspace*{10mm} Layer-1 & Layer-2 \\
                        \hline
                        TPX02&  3540&  1115& \hspace*{10mm}104& 123 \\
                        TPX12& -3540&  1146&  \hspace*{10mm}97& 113 \\
                         \hline\hline
\end{tabular}
\end{center}
\vspace*{-5mm}
\end{table}

\begin{figure}[tp]
\centering
\includegraphics[width=\linewidth]{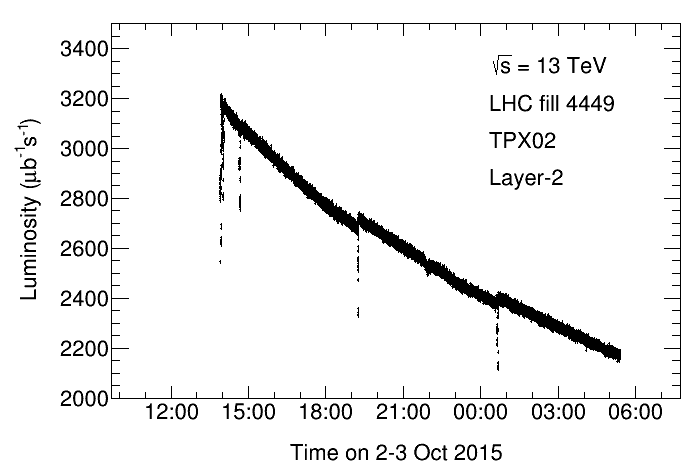}
\vspace*{-7mm}
\caption{Time history of the TPX luminosity.
The small dips, visible as variations from the descending curve, 
correspond to the times when the LHC operators performed small-amplitude 
beam-separation scans to optimize the luminosity.
The normalization between hit rate and luminosity is based on 
vdM scans using LHC fill 4266, detailed in Section~\ref{sec:vdM_scans}.}
\label{fig:tpxexample}
\vspace*{-3mm}
\end{figure}

Figure~\ref{fig:tpxexample} shows an example of the luminosity from hit-counting 
measured with TPX02 layer-2 for LHC fill 4449, taken on 2-3 October 2015.
All times are in GMT.

This article is structured as follows.
First, the concept of LHC luminosity monitoring from TPX hit-counting is introduced
in Section~\ref{sec:lumi_hit_counting}.
The vdM scan analysis based on hit-counting (Section~\ref{sec:vdM_scans}) is summarized
for an absolute luminosity calibration.
In Section~\ref{sec:lhcfit} the LHC luminosity curve is determined by
using the concept of averaged interactions per bunch-crossing,
together with the TPX measurement precision.
Section~\ref{sec:short-term_individual} describes the luminosity precision by evaluation of 
the difference between two layers of the same TPX device.
The long-term luminosity precision is given in Section~\ref{sec:longterm-term_individual}
from the comparison of layer-1 and layer-2 luminosity of the same TPX device.
The long-term luminosity precision from different TPX devices 
is given in Section~\ref{sec:longterm-term_distance}.
Finally, conclusions are given in Section~\ref{sec:conclusions}.

\section{LHC Luminosity from TPX Hit-Counting}
\label{sec:lumi_hit_counting}

The data from the TPX02 and TPX12 devices were used in hit-counting mode,
both having similar count rates as specified in Table~\ref{tab:tpx_detectors}.
The devices measure the luminosity independently and their measurements are 
cross-checked with each other.
A constant exposure time of 1\,s was used for the entire 2015 data-taking.

A small number of pixels becoming weak or noisy (e.g. due to radiation damage) 
could have a significant effect on the luminosity measurement.
Therefore, pixels with a count rate that is at least $3\sigma$ away from the mean, 
are excluded for each sensor region (uncovered and with converters). 
This requirement identifies about 7-13\% of the pixels on layer-1 and layer-2 
both for TPX02 and TPX12 per LHC fill, including 5-10\% of pixels at the boundaries 
of the sensor regions and edges of the sensor matrix.
Then, the logical OR of identified pixels per LHC fill was taken for all 2015 LHC fills to remove 
19-22\% of the total number of pixels of the sensors.
The effect of the pixel removal on the analysis was also studied with $2\sigma$ and $5\sigma$
criteria, with the result that the analysis
outcome regarding the LHC luminosity curve, measurement precision and long-term stability
remained unchanged.

The hit rate for the four TPX sensors is normalized to units of luminosity 
by multiplying with a scaling factor
as given in Section~\ref{sec:vdM_scans}. 

The induced radioactivity of material in the ATLAS cavern has 
no significant 
effect on the luminosity determination as determined by a dedicated study.

The ATLAS and CMS collaborations have elaborate systems of luminosity measurements,
described in~\cite{improvedLumiDet:2013},~\cite{Aaboud:2016hhf} (ATLAS) 
and~\cite{cms:mpx},~\cite{cms:2013} (CMS).
A comparative study of their results and the TPX luminosity monitoring
is beyond the scope of this article.

In addition to the hit-counting, luminosity can be measured with the
two other modes of TPX operation based on cluster-counting and summed energy deposits.
These luminosity measurements will be addressed in a separate publication.

The relation between the number of hits and clusters (particles) is investigated
in order to determine the statistical uncertainty of the luminosity measurement from hit-counting.
The average ratio of hits per cluster is approximately $R=N_{\rm hit}/N_{\rm cl}=10$, 
which was obtained with TPX
data from a low-intensity LHC fill for which the clusters on the sensors were well separated.
This factor is used for the statistical uncertainty determination in the following sections 
assuming that one cluster corresponds to one independent particle passing the 
device~\cite{Sopczak:2015qqj}.


\section{Van der Meer Scans}
\label{sec:vdM_scans}

Van der Meer (vdM) scans are used for absolute luminosity calibration at the 
LHC.
This scan technique was pioneered by Simon van der Meer 
at CERN in the 1960s~\cite{vdm} 
to determine the luminosity calibration in a simple way.
It involves scanning the LHC beams through one another to determine their sizes
in terms of the horizontal and vertical widths of the beams at the 
point of collision.
These width measurements are then combined with information of the number of circulating 
protons, allowing the determination of an absolute luminosity scale. 
The vdM scan analysis is based on the data taken on 24-25 August 2015
using TPX02 and TPX12 layer-1 and layer-2.

\begin{figure}[tp]
\vspace*{-2mm}
\centering
\includegraphics[width=\linewidth]{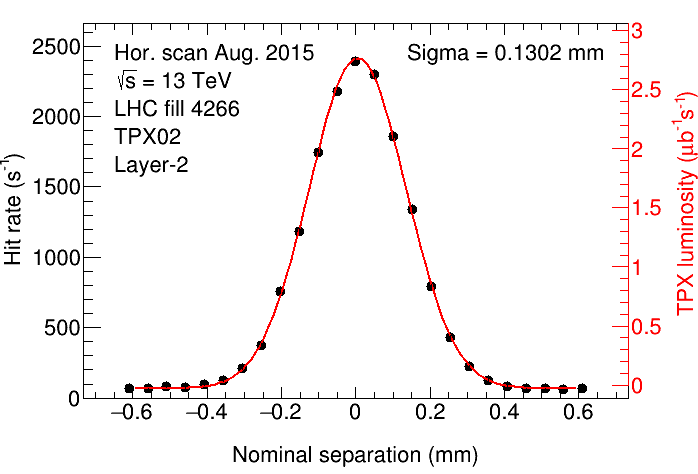}
\vspace*{-7mm}
\caption{Luminosity from hit-counting as a function of nominal beam separation
         measured with TPX02 layer-2
         during the first horizontal vdM scan in August 2015. 
Each data point shows the measured instantaneous luminosity 
before background subtraction.
Because the exposure time is significantly shorter than the duration of a 
scan step, the TPX samplings are averaged per scan step. 
The TPX samplings that partially or totally overlap with non-quiescent 
scan steps (varying beam separation) are not shown. 
The fit function is the sum of a single Gaussian 
(representing the proper luminosity in this scan) 
and a constant term that accounts for instrumental 
noise and single-beam background.
The TPX normalization uses this horizontal and a vertical 
beam width from LHC vdM fill 4266.
}
\label{fig:mpx01HitLumi2}
\vspace*{-6mm}
\end{figure}

The LHC beam separation dependence of the measured TPX luminosity is 
well represented by the sum of a single Gaussian and a constant 
(Fig.~\ref{fig:mpx01HitLumi2}).
The absolute luminosity normalization is derived 
from the combination of the hit rate, the horizontal and vertical 
convoluted widths and the average bunch currents.

\begin{figure}[hp]
\vspace*{-2mm}
\centering
\includegraphics[width=\linewidth]{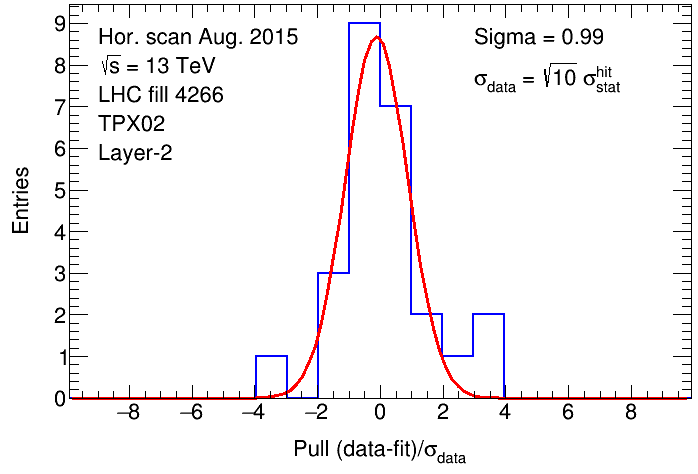}
\vspace*{-0.7cm}
\caption{Pull distribution defined as (data-fit)/$\sigma_{\rm data}$,
        where $\sigma_{\rm data} = \sqrt{R} \cdot \sigma_{\rm stat}^{\rm hit}$ 
        with $R=N_{\rm hit}/N_{\rm cl} =10$ for TPX02 layer-2.
        The data shown in Fig.~\ref{fig:mpx01HitLumi2} is used.
        LHC fill 4266.
}
\label{fig:vdmpull}
\vspace*{-4mm}
\end{figure}

The measurement uncertainty of the TPX devices can be determined with respect to 
the expected statistical uncertainty. For this study, the pull
distributions, as defined by (data-fit)/$\sigma_{\rm data}$, were determined, where  
$\sigma_{\rm data} = \sqrt{R} \cdot \sigma_{\rm stat}^{\rm hit}$, 
$\sigma_{\rm stat}^{\rm hit} = {\rm data}/\sqrt{N_{\rm hit}}$
and $R=10$.
Figure~\ref{fig:vdmpull} shows the pull distribution for the first horizontal vdM scan 
in August 2015, as seen by TPX02 layer-2.
The sigma of the pull distribution averaged over TPX02 and TPX12, layer-1 and layer-2
for both horizontal and vertical scans is $2.0\pm0.6$, which indicates that additional 
uncertainties are present beyond the determined statistical uncertainties, or 
correlations in the statistical evaluation have a significant effect as 
discussed in Ref.~\cite{Sopczak:2015qqj}. 
Furthermore, 
transverse proton-bunch profiles are not expected to be perfectly Gaussian; and even if they were,
a scan curve summed over Gaussian bunches of 
different widths would not be strictly Gaussian.  
Therefore, non-Gaussian contributions to 
the vdM-scan curves may contribute, 
at some level, 
to the 
widening of the pull distribution.

For TPX02 layer-2, 
the widths of the beam sizes (horizontal and vertical nominal beam separations) and 
their statistical uncertainties are $\Sigma_{\rm x}=(130.2\pm0.5) \,\micron$ and 
$\Sigma_{\rm y}=(118.6\pm0.5)\,\micron$, respectively.

The luminosity can be calculated as:                                                                     
\begin{equation}                                                                                         
L_{\rm TPX} = N_{\rm b} N_{\rm p1} N_{\rm p2} f/(2\pi \Sigma_{\rm x}\Sigma_{\rm y}),                     
\end{equation}                                                                                           
where                                                                                                    
$N_{\rm b}$ is the number of bunch crossings producing collisions per machine revolution,                
$N_{\rm p1}$ and $N_{\rm p2}$ are the average bunch populations (number of protons)                      
in beam~1 and beam~2, respectively,                                                                      
$f$ is the machine revolution frequency (11245.5~Hz), and                                                
$\Sigma_{\rm x}$ and $\Sigma_{\rm y}$ are the convoluted horizontal and vertical beam sizes.
The LHC parameters for fill 4266 are~\cite{lhc}:
\begin{itemize}
\item number of bunches: $N_{\rm b} =30$;
\item average number of protons (in units $10^{11}$) per bunch in beam~1 and in beam~2: 
      $N_{\rm p1}=26.5/30=0.883$ and $N_{\rm p2}=26.6/30=0.887$, respectively.
\end{itemize}
Thus, the resulting luminosity is 
$L_{\rm TPX} = 2.724\,{\rm \micro b^{-1}s^{-1}}$. 

The specific luminosity is defined as:
\begin{equation}
L_{\rm specific} =  L_{\rm TPX} / (N_{\rm b} N_{\rm p1} N_{\rm p2}) = f/(2\pi \Sigma_{\rm x}\Sigma_{\rm y}).
\end{equation}
Table~\ref{tab:vdm} summarizes the scan results for the first pair of vdM scans (horizontal and vertical)
registered with TPX02 and TPX12 for both their layers.

\begin{table}[htbp]
\vspace*{-1mm}
  \caption{Van der Meer (vdM) scan results for 2015 data.
           The scan was performed on 24-25 August (LHC fill 4266)
           and the first horizontal and vertical scans were used.
           The fit results for the first bunch-averaged horizontal $\Sigma_{\rm x}$ 
           and vertical $\Sigma_{\rm y}$ convoluted beam sizes are given,
           as well as the  specific luminosity.
           The hit rate at the peak 
           is averaged over the horizontal and vertical scans.
}
 \centering
\renewcommand{\arraystretch}{1.2} 
    \begin{tabular}{cccccc}
\hline\hline
TPX        & Layer     & $\Sigma_{\rm x}$ & $\Sigma_{\rm y}$ & $L_{\rm specific}$ & $N_{\rm peak}$ \\
                  &           & ($\micron$)   & ($\micron$)  
&\hspace*{-2mm}($\rm \micro b^{-1}s^{-1}/10^{25}$)\hspace*{-2mm} &  ($\rm hits/s$)\\ \hline
02         &  1      & 130.2 & 117.7 & 116.8 & 1495 \\
02         &  2      & 130.2 & 118.6 & 115.9 & 2386 \\
12         &  1      & 128.7 & 118.1 & 117.8 & 1386 \\
12         &  2      & 128.3 & 118.7 & 117.5 & 2298 \\
\hline\hline
    \end{tabular}%
  \label{tab:vdm}%
\vspace*{-2mm}
\end{table}%

For TPX02 layer 2, the fits of horizontal and vertical scans provide
$(2357\pm 14)$ and $(2415\pm 16)$ 
hits/s, respectively, at the peak above the background.
The average number is ($2386\pm11$) hits/s.
Thus, the normalization factor $n_{\rm f}$ between the TPX02 layer-2 hit 
rate and the instantaneous LHC luminosity is 
\begin{equation}
n_{\rm f}=\frac{\rm 2.724\,\micro b^{-1}s^{-1}}{\rm 2386\,hit\,s^{-1}}=
1.141\cdot 10^{-3}~{\rm \micro b^{-1} / hit}.
\end{equation}

The normalization factors for the other devices were calculated using the same procedure,
and the results are summarized in Table~\ref{tab:normalization}.

\begin{table}[tp]
\caption{Normalization ($1/n_{\rm f}$) to convert hit rates to luminosities.
         The larger values are for the thicker sensor layer.
\vspace*{-4mm}
}
\label{tab:normalization}
\begin{center}
\renewcommand{\arraystretch}{1.2}
             \begin{tabular}{ccc}
                        \hline\hline
                Device   & Layer & $1/n_{\rm f}$\\
                         &       & (hits/$\rm \micro b^{-1}$) \\\hline
                TPX02    & 1     &  544.8 \\
                TPX02    & 2     &  876.1 \\
                TPX12    & 1     &  500.9 \\
                TPX12    & 2     &  832.2 \\
\hline\hline
\end{tabular}
\end{center}
\vspace*{-9mm}
\end{table}

As already noted for the previous LHC Run-1 vdM scan analysis using MPX devices~\cite{Sopczak:2015qqj},
the normalization factor for the absolute luminosity is only approximate 
since the TPX exposure time is much longer than the bunch spacing.
Therefore, the bunch-integrated luminosity averages over the different bunch profiles.
In order to estimate the resulting uncertainty a simulation with 29 overlapping 
Gaussian distributions was performed~\cite{Sopczak:2015qqj}, which led to an estimate of the 
resulting uncertainty on the normalization factor (from this source only) of about 1\%.

Although further uncertainties could arise from non-Gaussian shapes, this study shows that the 
Gaussian approximation of the sum of Gaussians is quite robust with the TPX system 
and the luminosity approximation by bunch integration is a sensible approach.
No attempt was made for a precise determination of the total uncertainty which would require a
dedicated study~\cite{improvedLumiDet:2013,Aaboud:2016hhf}.

\section{LHC Luminosity Curve and TPX Precision}
\label{sec:lhcfit}
The TPX network has the capability to study the LHC luminosity curve with precision.
Six LHC fills of proton-proton collisions were investigated in detail.
As an example, details are given for the LHC fill 4449, taken on 3 October 2015.

First, the TPX luminosity is calculated using the normalization factors from Table~\ref{tab:normalization}
and it is then converted to an average number of inelastic interactions per bunch crossing by
\begin{equation}
\mu=L\cdot \sigma_{\rm inel}/(N_{\rm b}\cdot f), 
\end{equation}
where $N_{\rm b}=1453$ colliding bunches, $f=11245.5$~Hz and the inelastic cross-section 
$\sigma_{\rm inel} = 80$\,mb~\cite{inelastic}.
Accelerator simulations~\cite{Antoniou:2016mwq} have shown that under routine physics conditions ($\beta^* < 1$\,m),  
elastic proton-proton scattering contributes negligibly to the particle-loss rate.
This is because the typical scattering angle is so small compared to the natural angular 
divergence of the beam that the protons remain 
with the dynamic aperture of the ring.

\vspace*{-2mm}
\subsection{Fitted LHC Luminosity Curve}
In a simple approximation,
the loss rate of protons $N$ in the colliding beam is governed by:
\begin{equation}
-dN/dt = \lambda_{\rm bb}N^2/N_0+\lambda_{\rm g}N,
\label{eq:rate}
\end{equation}
where $N_0$ is the initial number of protons, and $\lambda_{\rm bb}$ and $\lambda_{\rm g}$ 
are constants related to beam-beam (burning-off the proton bunches) and 
single bunch (e.g. beam-gas) interactions, respectively.
This equation has a known solution, which is used as fit function:
\begin{equation}
N(t) = \frac{N_0 {\rm e}^{-\lambda_{\rm g}t}}{1+ \frac{\lambda_{\rm bb}}{\lambda_{\rm g}}(1-{\rm e}^{-\lambda_{\rm g}t})},
\label{eq:exponent}
\end{equation}
with two well-known border cases:
\begin{equation}
N(t) = N_0 {\rm e}^{-\lambda_{\rm g}t}~ {\rm for}~ \lambda_{\rm bb}\ll \lambda_{\rm g}~{\rm and},
\label{eq:exponenta}
\end{equation}
\begin{equation}
N(t) = \frac{N_0}{1+ \lambda_{\rm bb} t}~{\rm for}~\lambda_{\rm g}\ll \lambda_{\rm bb}.
\label{eq:exponentb}
\end{equation}

Next, the expected mean lifetime of inelastic beam-beam interactions is calculated
from the LHC beam parameters, as given in Table~\ref{tab:2015LHC}, and used as a fixed parameter to
determine $\lambda_{\rm g}$. A simultaneous fit of 
$\lambda_{\rm bb}$ and $\lambda_{\rm g}$ did not converge.
As already noted in the previous LHC Run-1 analysis~\cite{Sopczak:2015qqj}, 
$\lambda_{\rm bb}$ and $\lambda_{\rm g}$ are strongly anti-correlated.
Compared to the data-taking at Run-1, the Run-2 LHC luminosity curve is flatter, making the fit
less sensitive to the fit parameters. 

The mean lifetime from inelastic beam-beam interactions is given by~\cite{LHC94}:
\begin{equation}
\label{eq:tbb}
t_{\rm bb}^{\rm inel} = N_{\rm b}N_0/(N_{\rm exp}L_0 \sigma_{\rm inel}),
\end{equation}
where 
$N_0$ is the initial number of protons per bunch 
($N_{\rm b}N_0 = 1.6\cdot 10^{14}$ protons~\cite{lhc}).
For LHC fill 4440 the initial luminosity is 
$L_0 = 3360\,{\rm \micro b^{-1} s^{-1}}$~\cite{lhc},
the number of experiments is $N_{\rm exp} = 2$ (ATLAS~\cite{atlasCollaboration:2013} 
and CMS~\cite{cmsCollaboration}).
For the lifetime $t_{\rm bb}^{\rm inel} = 3.07\cdot 10^5\,{\rm s}$ is obtained, thus 
\begin{equation}
\lambda_{\rm bb}^{\rm inel} = 1/t_{\rm bb}^{\rm inel} = 0.2815\,{\rm day}^{-1}.
\end{equation}

The value $\lambda_{\rm bb}^{\rm inel}$ depends 
on the initial luminosity and the initial number of protons,
thus on the starting value of $\mu_1=10.43$ for the first fit. 
Since $L \propto N^2$,
one can write $ \lambda_{\rm bb}^{\rm inel}  \propto \sqrt{L_0} \propto \sqrt{\mu_0} $. 
Thus for the lower initial luminosity in the fit, a longer lifetime is expected from
beam-beam interactions and therefore a smaller 
\begin{equation}
\lambda^{1}_{\rm bb} =\sqrt{10.51/24.8}\cdot 0.2815\,{\rm day}^{-1}= 0.1833  \,{\rm day}^{-1}.
\label{eq:inel}
\end{equation}

The frequent LHC small-amplitude beam-separation scans for
optimisation of the luminosity
made it necessary to adapt the fit function.
Therefore, the data during the scans is removed and the LHC luminosity curve is fitted 
with a function having the values of $\lambda_{\rm bb}$ reduced by $\sqrt{\mu_i/\mu_0}$ 
for each time period between the scans.
The $\mu_i$ values are calculated for the starting values for each region between the scans,
and $\mu_0$ is the value for the start of the LHC fill.

\begin{figure*}[htbp]
\centering
\includegraphics[width=0.49\linewidth]{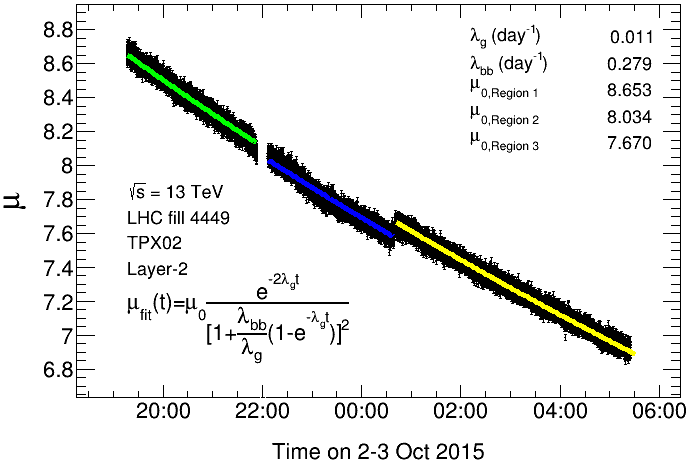}
\includegraphics[width=0.49\linewidth]{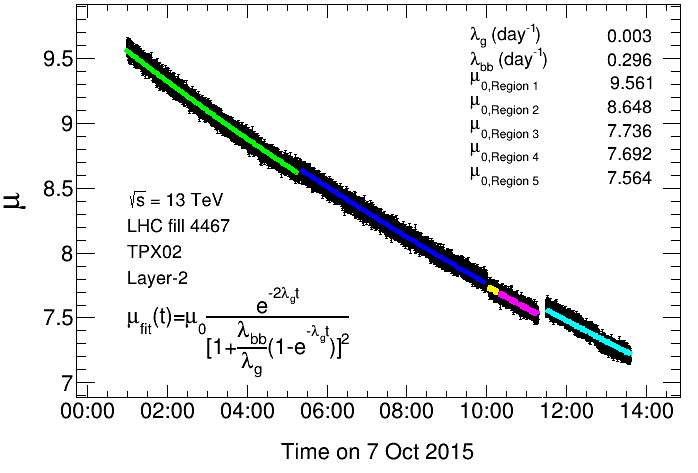}
\vspace*{-3mm}
\caption{Average number of interactions per bunch crossing  
         as a function of time, seen by TPX02 layer-2 in hit-counting mode.
         The distribution is approximately described by a function as 
         given in the figure.
         The parameters are defined in the text.
         The statistical uncertainties per data point are indicated.
         They depend on the hit statistics scaled by a factor $\sqrt{10}$.
         For the first time intervals the $\lambda_{\rm bb}$ is calculated analytically
         from the LHC parameters, and for the subsequent time intervals 
         $\lambda_{\rm bb}$ is scaled with $\sqrt{\mu_i/\mu_1}$, where the index
         corresponds to the time period.
}
\label{fig:fit02} 
\vspace*{-4mm}
\end{figure*}

Six long LHC fills with large luminosity are investigated as given in Table~\ref{tab:2015LHC}.
Figure~\ref{fig:fit02} shows the fit of TPX02 layer-2 data for two of these LHC fills.

The fit results indicate that the LHC luminosity reduction is predominantly reduced by  
beam-beam interactions since a larger value of $\lambda_{\rm bb}$ corresponds to a shorter lifetime. 
In order to determine the $\lambda_{\rm g}$ and its uncertainty, the fit is performed
for the six LHC fills and the four sensors separately, resulting in a large single-bunch 
lifetime $1/\lambda_{\rm g}$: 
\begin{equation}
\rm \lambda_{g}^{avg} = (0.04\pm0.03)~ day^{-1},
\end{equation}
where the average is taken over 24 measurements, 
and the uncertainty is given as the square root of the variance.

\vspace*{-2mm}
\subsection{Precision}
In order to investigate the precision of the TPX luminosity measurements,
the difference between the fit and the data is studied as a function of time.
As an example, the data is analysed from the last continuous LHC curve
taken on 3 October 2015 for about 4.5 hours starting 0:45.

A precision of 0.35\% has been obtained with TPX02 layer-2 
and similarly with other TPX sensors, as listed in Table~\ref{tab:precision}.
The given uncertainties result from statistical and systematic uncertainties by the TPX 
measurements convoluted with uncertainties arising from fluctuations in the 
proton-proton collision rates.

\begin{figure*}[htbp]
\centering
\includegraphics[width=0.49\linewidth]{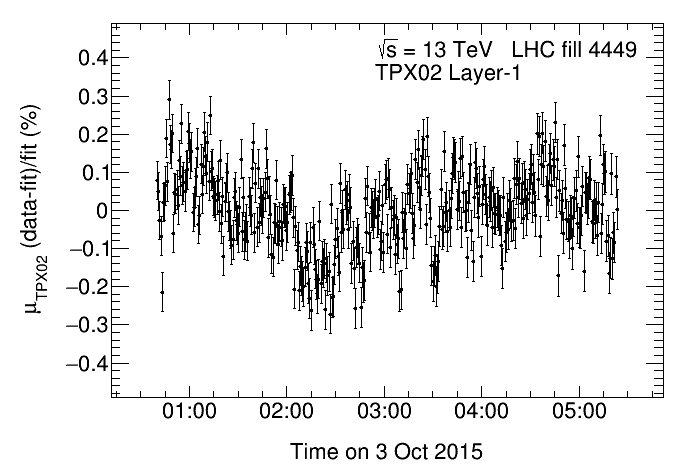}
\includegraphics[width=0.49\linewidth]{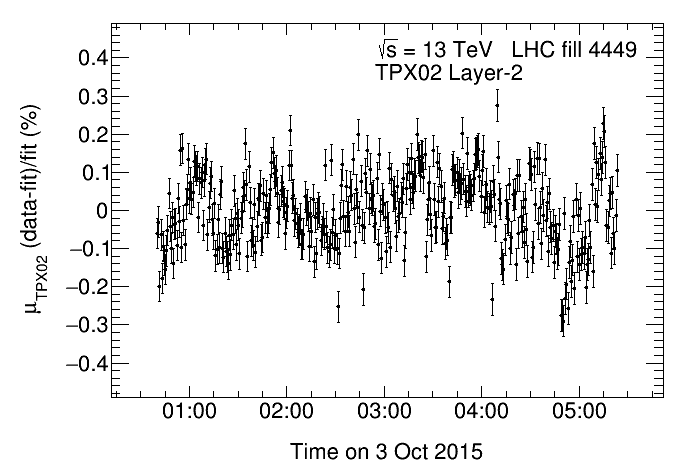}
\includegraphics[width=0.49\linewidth]{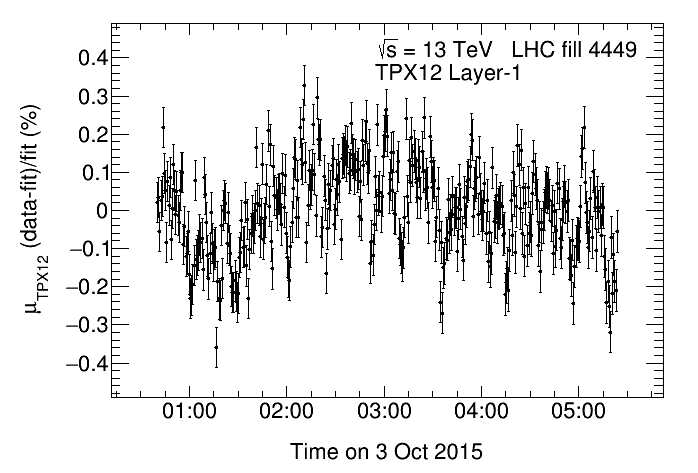}
\includegraphics[width=0.49\linewidth]{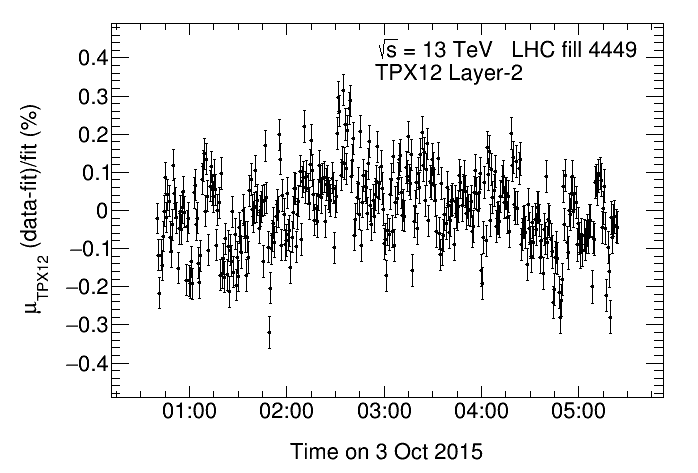}
\vspace*{-0.3cm}
\caption{Relative difference between data and fitted average 
         number of interactions per bunch crossing as a function
         of time, seen by TPX02 and TPX12 for layer-1 and layer-2.
         The data is shown for 30 frames (1\,s each) combined
         in order to decrease the statistical uncertainty.
         LHC fill 4449.
}
\label{fig:residual_30frames}
\end{figure*}

It is noted that the TPX measurement precision is statistically limited by the number of hits per frame. 
Therefore, 10, 20, 30 and 40 frames are grouped together. 
Consequently, the statistical precision significantly increases.
Figure~\ref{fig:residual_30frames} shows the residuals 
for groups of 30 frames.
The four distributions show no particular uniform structure and therefore no LHC luminosity
variation at the 0.1\% level.
Table~\ref{tab:precision} lists the obtained precisions for TPX02 and TPX12 layer-1 and layer-2.
The fit results using the data of the four sensors are consistent.

\begin{table*}[hbp]
\caption{Parameters of the LHC proton-proton collisions for the 2015 fills analysed 
         with TPX02 and TPX12 in the fitting of LHC luminosity reduction curve.
         $L_0$ is the luminosity and $\mu_0$ is the average number of interactions per bunch crossing
         at the start of the LHC fill. The $\lambda_{\rm bb}^0$ is given for the start of the LHC fill,
         and $\lambda_{\rm bb}^1$ is given at the start time of the fit for the first time period.
\vspace*{-3mm}
}
\label{tab:2015LHC}
\begin{center}
\renewcommand{\arraystretch}{1.2}
             \begin{tabular}{ccccccccc}
                        \hline\hline
 Date    & Fill   & Start time & $kN_0$      & $L_0$ & $\mu_0$ & Colliding & $\lambda_{\rm bb}^0$ & $\lambda_{\rm bb}^1$ \\
in 2015 & & (unix time)  & $(10^{11})$ & $(\rm \micro b^{-1}s^{-1})$ &      & bunches   & $\rm (day^{-1})$     &$\rm (day^{-1})$ \\ \hline
29 Sep. & 4440 & 1443525360 & 1650 & 3360 & 24.8 & 1453 & 0.2815 & 0.1833 \\
2-3 Oct.  & 4449 & 1443785520 & 1607 & 3240 & 26.0 & 1453 & 0.2787 & 0.1615 \\
6-7 Oct.  & 4467 & 1444149360 & 1729 & 3700 & 24.1 & 1596 & 0.2958 & 0.1873 \\
9-11 Oct.  & 4479 & 1444427640 & 1950 & 4240 & 24.6 & 1813 & 0.3006 & 0.2194 \\
30-31 Oct. & 4557 & 1446221040 & 2388 & 4450 & 24.7 & 2232 & 0.2576 & 0.1840 \\
31 Oct.-1 Nov.  & 4560 & 1446311100 & 2610 & 5020 & 24.5 & 2232 & 0.2659 & 0.1859 \\
 \hline \hline
\end{tabular}
\end{center}
\end{table*}

\begin{table*}[htbp]
\caption{Precision of TPX02 and TPX12 luminosity measurements for the average of layer-1 and layer-2.
         The precision is  given for 1 frame and for 10, 20, 30 and 40 frames 
         combined.
         The width of the Gaussian fit $\sigma_{\rm (data-fit)/fit}$ gives the precision of the 
         measurement and $\sigma_{\rm pull}$ gives the width of the fit of the pull distribution
         with statistical uncertainties only.
         LHC fill 4449.
\vspace*{-3mm}
}
\label{tab:precision}
\begin{center}
\renewcommand{\arraystretch}{1.2}
             \begin{tabular}{ccccccccc}
                        \hline\hline
Frames&\multicolumn{4}{c}{$\sigma_{\rm (data-fit)/fit}$ (\%)} &\multicolumn{4}{c}{$\sigma_{\rm pull}$} \\
used  &\multicolumn{2}{c}{TPX02} &\multicolumn{2}{c}{TPX12} 
      &\multicolumn{2}{c}{TPX02} &\multicolumn{2}{c}{TPX12} \\
      &Layer-1 &Layer-2 &Layer-1 &Layer-2 &Layer-1 &Layer-2 &Layer-1 &Layer-2 \\ \hline
   1  & 0.36 & 0.35 & 0.37 & 0.36 & 1.27 &1.58 & 1.28 & 1.57\\
   10 & 0.13 & 0.13 & 0.15 & 0.13 & 1.47 &1.87 & 1.61 & 1.82\\
   20 & 0.11 & 0.10 & 0.13 & 0.11 & 1.71 &2.04 & 1.96 & 2.12\\
   30 & 0.10 & 0.09 & 0.11 & 0.10 & 1.96 &2.18 & 2.17 & 2.40\\
   40 & 0.09 & 0.08 & 0.11 & 0.09 & 2.13 &2.39 & 2.24 & 2.56\\
                \hline\hline
\end{tabular}
\end{center}
\end{table*}

\begin{figure}[hp]
\centering
\includegraphics[width=\linewidth]{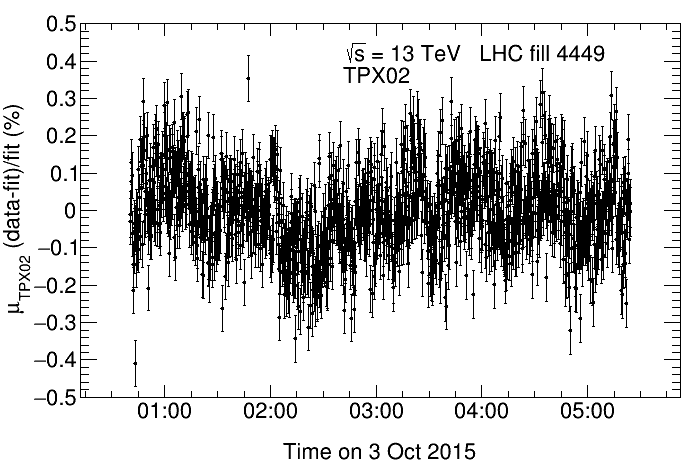}
\vspace*{-8mm} 
 \caption{Relative difference between data and fitted average 
         number of interactions per bunch crossing as a function
         of time, seen by TPX02 averaged over layer-1 and layer-2.
         The data is shown for 10 frames (1\,s each) combined
         in order to decrease the statistical uncertainty.
         LHC fill 4449.}
\label{fig:precision}
\vspace*{5mm} 
\end{figure}

\begin{figure}[hp]
\centering
\includegraphics[width=\linewidth]{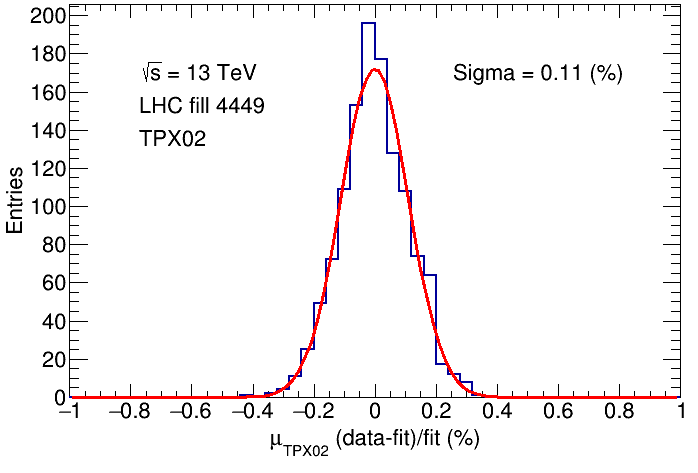}
\vspace*{-8mm} 
 \caption{Gaussian fit of the relative difference between data and fitted average 
         number of interactions per bunch crossing as a function
         of time, seen by TPX02 averaged over layer-1 and layer-2.
         The data is shown for 10 frames (1\,s each) combined
         in order to decrease the statistical uncertainty.
         LHC fill 4449.}
\label{fig:precision_gauss}
\end{figure}

\begin{figure}[t]
\centering
\includegraphics[width=\linewidth]{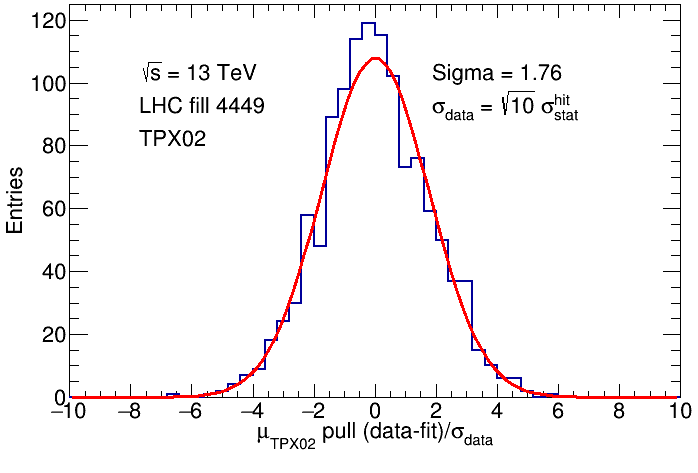}
\vspace*{-8mm} 
\caption{Pull distribution defined as (data-fit)/$\sigma_{\rm data}$,
        where $\sigma_{\rm data} = \sqrt{R} \cdot \sigma_{\rm stat}^{\rm hit}$.
        with $R=N_{\rm hit}/N_{\rm cl}=10$.
        The data is averaged over layer-1 and layer-2 for 
        TPX02 and TPX12 with 10 frames (1\,s each) combined.
        The data shown in Fig.~\ref{fig:precision_gauss} is used.
        LHC fill 4449.
}
\label{fig:precision_pull}
\end{figure}

In order to increase the statistical significance of each TPX device the luminosity
measured by layer-1 and layer-2 is averaged, and presented in Fig.~\ref{fig:precision} 
for 10 frames combined.
It shows the relative difference between data and fitted average number of interactions 
per bunch crossing as a function of time, seen by TPX02.
The corresponding Gaussian fit is shown in Fig.~\ref{fig:precision_gauss}.
Figure~\ref{fig:precision_pull} shows the pull distribution assuming statistical 
uncertainties only.
Table~\ref{tab:precision_average} lists the obtained uncertainties and the corresponding 
pull values for 1\, frame, as well as for 10, 20, 30 and 40 frames combined.
The resulting relative short-term luminosity measurement precision is 0.1\% for 10\,s time intervals.
As a consistency check of the obtained precision,
the data from 7 October 2015 (LHC fill 4467) from 5:30 to 10:00 was analysed in the same way, 
reproducing the results on the precision.
Thus, the resulting luminosity precision of the TPX system is higher compared to the previous MPX 
luminosity measurement precision (0.3\% for 60\,s time interval)~\cite{Sopczak:2015qqj} for data
taken at LHC during Run-1 operation.

\begin{table}[hp]
\caption{Precision of TPX02 and TPX12 luminosity measurements with respect to the fitted curve
         for the average of layer-1 and layer-2.
         The precision is  given for 1 frame and for 10, 20, 30 and 40 frames 
         combined.
         The width of the Gaussian fit $\sigma_{\rm (data-fit)/fit}$ gives the precision of the 
         measurement and $\sigma_{\rm pull}$ gives the width of the fit of the pull distribution
         with statistical uncertainties only.
         LHC fill 4449.
}
\label{tab:precision_average}
\vspace*{-2mm}
\begin{center}
\renewcommand{\arraystretch}{1.2}
             \begin{tabular}{ccccc}
                        \hline\hline
                Frames &\multicolumn{2}{c}{ $\sigma_{\rm (data-fit)/fit}$} (\%) 
                       &\multicolumn{2}{c}{ $\sigma_{\rm pull}$} \\
                used   & TPX02 & TPX12 &  TPX02 & TPX12   \\ \hline
                1  & 0.30 & 0.30
& 1.50 & 1.57\\
                10 & 0.11 & 0.12
& 1.76 & 1.94\\
                20 & 0.09 & 0.10
& 1.96 & 2.36\\
                30 & 0.08  & 0.09
& 2.21 & 2.56 \\
                40 & 0.08  & 0.09
& 2.35 & 2.88\\
\hline\hline
\end{tabular}
\end{center}
\vspace*{-5mm}
\end{table}

\section{Short-term Precision of Individual TPX Devices}
\label{sec:short-term_individual}
\begin{figure}[tp]
\centering
\includegraphics[width=\linewidth]{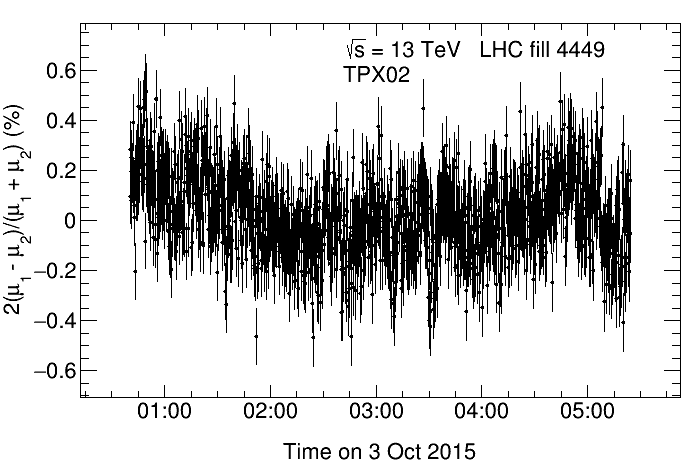}
\vspace*{-7mm} 
\caption{Relative difference between the average number of interactions per bunch 
        crossing measured by layer-1 and layer-2 of TPX02 as a function
        of time.
         The data is shown for 10 frames (1\,s each) combined
         in order to decrease the statistical uncertainty.
LHC fill 4449.}
\label{fig:shortterm_layer}
\end{figure}

\begin{figure}[hp]
\centering
\includegraphics[width=\linewidth]{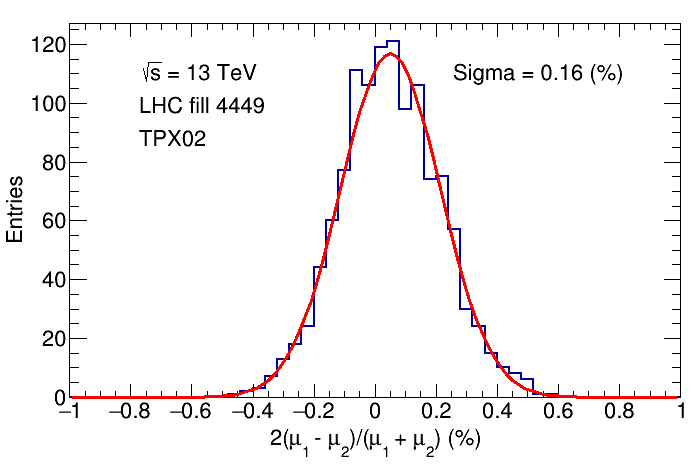}
\vspace*{-8mm} 
\caption{Gaussian fit of the relative difference between the average number of interactions per bunch 
        crossing measured by layer-1 and layer-2 of TPX02 and TPX12 as a function of time.
         The data is shown for 10 frames (1\,s each) combined
         in order to decrease the statistical uncertainty.
LHC fill 4449.}
\label{fig:shortterm_layer_gauss}
\end{figure}

\begin{figure}[bp]
\centering
\includegraphics[width=\linewidth]{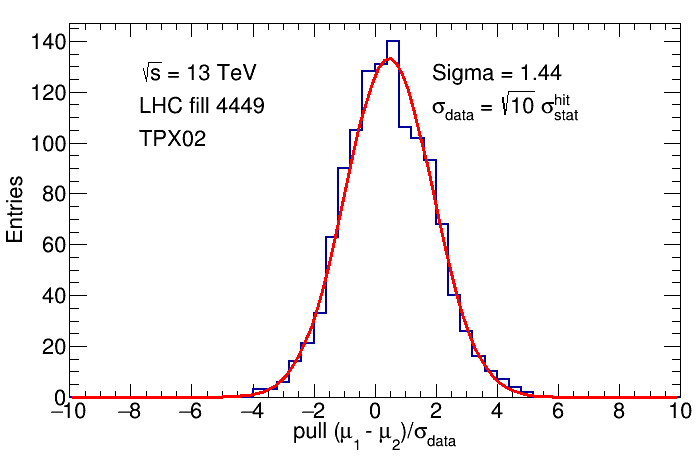}
\vspace*{-8mm} 
\caption{Pull distribution defined as $(\mu_2-\mu_1)/\sigma_{\rm data}$,
        where $\sigma_{\rm data} = \sqrt{R} \cdot \sigma_{\rm stat}^{\rm hit}$
        with $R=N_{\rm hit}/N_{\rm cl}=10$ for TPX02.
        The data shown in Fig.~\ref{fig:shortterm_layer_gauss} is used.
       LHC fill 4449.
}
\label{fig:shortterm_layer_pull}
\end{figure}

In order to determine the short-term precision of individual TPX devices for luminosity measurements,
the relative difference in luminosity measured by layer-1 and layer-2 of the same 
TPX device is studied as a function of time.
The statistical precision is increased by grouping
10, 20, 30 and 40 frames.
Figure~\ref{fig:shortterm_layer}  shows the fit results for 10 frames combined, 
while Fig.~\ref{fig:shortterm_layer_gauss} shows the corresponding precision
as the width of the Gaussian fit.
The resulting pull distribution is shown in Fig.~\ref{fig:shortterm_layer_pull}.
For single frames the width of the pull distribution is about unity, indicating that the
statistical uncertainty is dominant.
The width of the pull distribution is increasing as more frames are combined in order to 
decrease the statistical uncertainty, and the pull values indicate that in addition
systematic uncertainties are present.

Table~\ref{tab:shortterm_layer} lists the obtained precisions for TPX02 and TPX12.
As the difference of two measurements (from layer-1 and layer-2) is calculated
and the statistical significance of each measurement is about the same, the uncertainty of each 
measurement is about $\sqrt{2}$ of overall uncertainty, thus leading to a measurement precision
for each TPX device of approximately 0.1\% for 10\,s time intervals.

\begin{table}[hp]
  \caption{
         Precision of TPX02 and TPX12 luminosity measurements between layer-1 and layer-2.
         The precision is  given for 1 frame and for 10, 20, 30 and 40 frames 
         combined.
         The width of the Gaussian fit $\sigma_{\rm 2(\mu_1-\mu_2)/(\mu_1+\mu_2)}$  
         gives the precision of the 
         measurement and $\sigma_{\rm pull}$ gives the width of the fit of the pull distribution
         with statistical uncertainties only.
         LHC fill 4449.
}
 \centering
\renewcommand{\arraystretch}{1.2} 
    \begin{tabular}{ccccc}
\hline\hline
                Frames &\multicolumn{2}{c}{$\sigma_{\rm 2(\mu_1-\mu_2)/(\mu_1+\mu_2)}$ (\%)} 
                       &\multicolumn{2}{c}{$\sigma_{\rm pull}$} \\
                used &TPX02&TPX12&TPX02&TPX12 \\ \hline
                1  & 0.41 & 0.41 & 1.15 & 1.12\\
                10 & 0.16 & 0.17 & 1.44 & 1.45\\
                20 & 0.14 & 0.14 & 1.66 & 1.66\\
                30 & 0.12 & 0.13 & 1.89 & 2.04\\
                40 & 0.11 & 0.13 & 1.95 & 2.09\\
\hline\hline
\end{tabular}
\label{tab:shortterm_layer}
\end{table}

\clearpage
\section{Long-term Stability of Individual TPX Devices}
\label{sec:longterm-term_individual}
\begin{figure}[hbtp]
\begin{minipage}[b]{2.0\linewidth}
\vspace*{-7mm}
\centering
\includegraphics[width=0.49\linewidth]{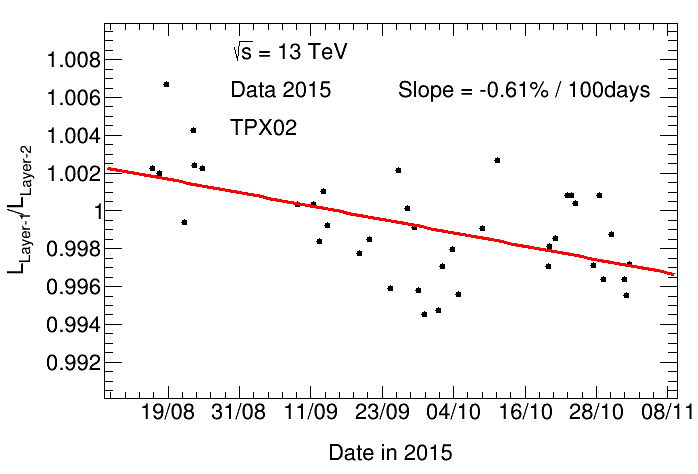}
\includegraphics[width=0.49\linewidth]{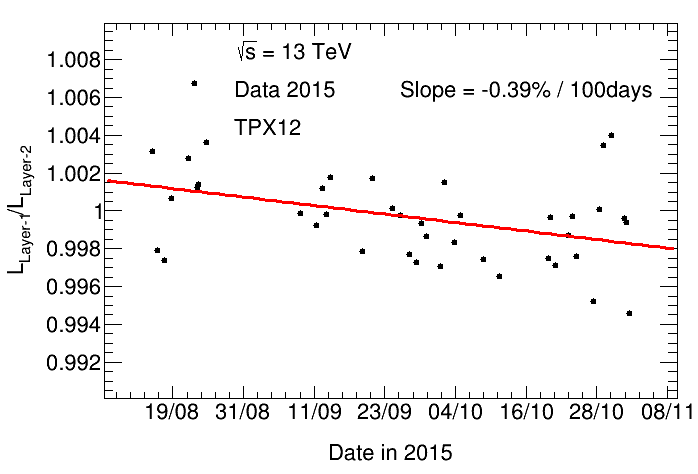}
\vspace*{-3mm}
\caption{Time history of the luminosity ratio measured by layer-1 and layer-2 for TPX02 and TPX12.
         The 2015 TPX data is divided into luminosity blocks of typically one minute length and 
         then grouped into LHC fill time periods.
         The size of the statistical error bar is below the size of the data point.
         A linear fit is applied to determine the slope.
         LHC fills from August to November 2015.
}
\label{fig:internal_hit14}
\vspace*{-5mm}
\end{minipage}
\end{figure}

The long-term time stability of the luminosity monitoring is determined for individual TPX devices
by comparing the luminosity measured by the two separate sensitive layers of TPX02 and TPX12.
For this analysis the frames are grouped corresponding to time periods of the 2015 LHC fills
with instantaneous luminosity above $\rm 1000\,\micro b^{-1}s^{-1}$.
A linear fit is applied to the (TPX layer-1)/(TPX layer-2) luminosity ratio versus time 
for the August to November 2015 data-taking period, as given in Fig.~\ref{fig:internal_hit14}.
The slope of the linear fit is taken as a measure of time stability.
The uncertainty is obtained from the fit. 
As the statistical uncertainty of the 
ratio measurements for the grouped frames is much less than the systematic uncertainties,
each ratio is given equal weight in the fit.
The obtained slope values and their uncertainties for TPX02 and TPX12 are summarized in 
Table~\ref{tab:MPX_SUM_used14}.
The slope shows that the luminosity ratio measured by layer-1 and layer-2
slightly decreases with time. This could indicate a small relative change in sensitivity with
time between the thinner and thicker sensors.
The time-stability of the luminosity measurements between individual layers is  
about 0.5\% per 100 days.

In addition, a study of the internal calibration transfer was performed.
As described in Section~\ref{sec:vdM_scans} the normalizations of the TPX sensors were performed with 
a vdM scan using LHC fill 4266 (24-25 August 2015) with a peak luminosity of 
$\rm 2.7\,\micro b^{-1}s^{-1}$.
In the same LHC fill, the luminosity was kept constant after the vertical and horizontal scans 
for about 5 hours.
Using the data of this time period the \mbox{(TPX layer-1)/}(TPX layer-2) luminosity ratio 
was determined with TPX02 and TPX12, 
and found to be in agreement with the ratios at high luminosity within 0.5\% uncertainty.

\begin{table}[hbp]
\caption{\label{tab:MPX_SUM_used14}
         Slope of time history of the luminosity ratio measured by layer-1 and layer-2 
for TPX02 and TPX12.
The slope values and the uncertainties are given per second and in percent per 100 days. 
}
\begin{center}
\renewcommand{\arraystretch}{1.2}
\begin{tabular}{ccccc}
\hline\hline
TPX  & Slope       & $\sigma$Slope    & Slope     & $\sigma$Slope \\
     & ($10^{-10}\,{\rm s}^{-1}$)&($10^{-10}\,{\rm s}^{-1}$) & (\%/100d) & (\%/100d)   \\ \hline
02   &   $-7.06$    &1.71  &   $-0.61$    &  0.15 \\
12   &   $-4.51$    &1.62  &   $-0.39$    &  0.14 \\
\hline\hline
\end{tabular}
\end{center}
\end{table}

\section{Long-term Stability of Different TPX Devices}
\label{sec:longterm-term_distance}
\begin{figure}[tp]
\centering
\includegraphics[width=\linewidth]{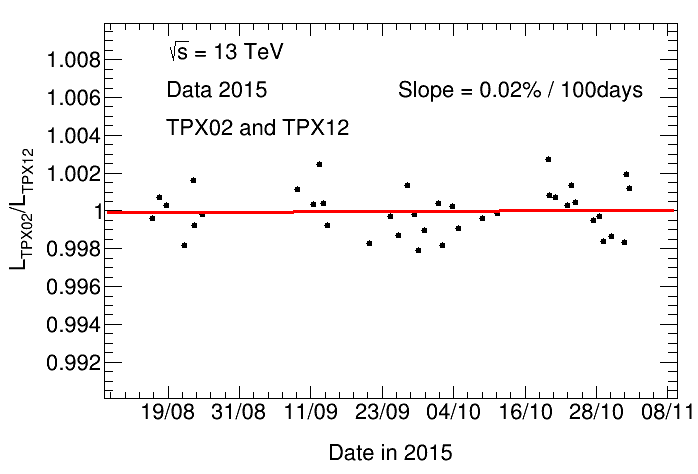}
\vspace*{-8mm}
\caption{Time history of the luminosity ratio measured by the TPX02 and TPX12 devices averaged 
         over the data taken by layer-1 and layer-2.
         The 2015 TPX data is divided into luminosity blocks of typically one minute length and 
         then grouped into LHC fill time periods.
         The size of the statistical error bar is below the size of the data point.
         A linear fit is applied to determine the slope.
         LHC fills from August to November 2015.
}
\label{fig:internal_hit14LB}
\vspace*{-3mm}
\end{figure}

The long-term time stability of the luminosity monitoring is determined for two TPX devices
located about 7\,m apart at opposite sides of the primary proton-proton interaction point.
First, the frames 
are grouped into luminosity blocks of about one minute.
This grouping of frames is necessary for a comparative study as the 
start times of the frames are not synchronized between the TPX devices.
Then, the luminosity blocks are grouped into time periods corresponding to the LHC fills
with instantaneous luminosity above $\rm 1000\,\micro b^{-1}s^{-1}$.
The luminosity ratio measured by TPX02 and TPX12
is calculated and a linear fit is applied for the August to November 2015 data-taking period,
as given in Fig.~\ref{fig:internal_hit14LB}.
The slope of the linear fit is taken as a measure of time stability.
The obtained slope value and its uncertainty is summarized in Table~\ref{tab:MPX_SUM_used14_2}.
The uncertainty is obtained from the fit. 
As the statistical uncertainty of the 
ratio measurements for the grouped time intervals is much less than the systematic uncertainties,
each ratio is given equal weight in the fit.
The fluctuations are much larger than the slope of 0.02\% per 100 days,
therefore, the largest fluctuations are used as an estimate of the long-term time stability.
These fluctuation could either result from the TPX operation, or from small variations in the 
complex LHC radiation field depending on small changes in the colliding beam optics.
Thus, conservatively, the study of two different TPX devices indicates an internal time stability of the 
luminosity measurement below 0.5\% per 100 days.

\begin{table}[hp]
\caption{\label{tab:MPX_SUM_used14_2}
         Slope of time history of the luminosity ratio measured by TPX02 and TPX12 
         averaged over the data recorded by layer-1 and layer-2.
         The slope values and the uncertainties are given per second 
         and in percent per 100 days.
}
\begin{center}
\renewcommand{\arraystretch}{1.2}
\begin{tabular}{ccccc}
\hline\hline
TPX  & Slope       & $\sigma$Slope    & Slope     & $\sigma$Slope \\
     & ($10^{-10}\,{\rm s}^{-1}$) &($10^{-10}\,{\rm s}^{-1}$) &(\%/100d) & (\%/100d)  \\ \hline
02/12        &   0.25    &0.89  &   0.02   &  0.08   \\
\hline\hline
\end{tabular}
\end{center}
\end{table}

\clearpage

\newpage
\section{Conclusions}
\label{sec:conclusions}

The network of TPX devices installed in the ATLAS detector cavern has successfully 
taken data at LHC during Run-2 with 13 TeV proton-proton collisions.
An approximate absolute luminosity calibration was determined from a vdM scan in 2015.
The TPX network measured the LHC luminosity curve with precision indicating that 
the luminosity reduction from single-bunch interactions was much less than from beam-beam interactions.
The relative short-term precision of the TPX luminosity measurements was determined to be 
0.1\% for 10\,s time intervals,
and the internal long-term stability of the TPX system for luminosity measurements was below
0.5\% for the 2015 data-taking period.

\section*{Acknowledgment}
The authors would like to thank warmly the ATLAS Luminosity Group for useful discussions and interactions,
and the Medipix Collaboration for providing the TPX assemblies.
The project is supported by the 
Ministry of Education, Youth and Sports of the Czech Republic under projects number 
LG 15052 and LM 2015058, and the Natural Sciences and 
Engineering Research Council of Canada (NSERC).
Calibration measurements were performed at the Prague Van-de-Graaff accelerator 
funded by the Ministry of Education, Youth and Sports of the Czech Republic
under project number LM 2015077.
\bibliographystyle{IEEEtran}

\newpage
\bibliography{biblio}

\end{document}